\documentclass[12pt]{iopart}
\usepackage{iopams}
\usepackage[numbers,sort&compress]{natbib}
\usepackage{graphicx}
\usepackage{dcolumn}
\usepackage{bm}
\usepackage{color}

\renewcommand{\vec}[1]{\mathbf{#1}}

\begin{document}
\title{Ice Rule Breakdown and frustrated antiferrotoroidicity in an
artificial colloidal Cairo ice}
\author{Carolina Rodr\'iguez-Gallo$^{1,2}$, Antonio Ortiz-Ambriz$^{1,3}$,
Cristiano Nisoli$^{4}$, and Pietro Tierno$^{1,2,5}$}
\address{
$^{1}$Departament de F\'isica de la Mat\`eria Condensada, Universitat de Barcelona, 08028, Barcelona, Spain.\\
$^{2}$Universitat de Barcelona Institute of Complex Systems (UBICS), Universitat de Barcelona, 08028 Spain.\\
$^{3}$Tecnologico de Monterrey, Escuela de Ingenier\'{i}a y Ciencias, Campus Monterrey, 64849 Mexico.\\
$^{4}$Los Alamos National Laboratory, Los Alamos, NM 87545, United States of
America.\\
$^{5}$Institut de Nanoci\`{e}ncia i Nanotecnologia, Universitat de Barcelona, 08028 Spain.}
\ead{ptierno@ub.edu}
\vspace{2pc}
\maketitle
\begin{abstract}
We combine experiments and numerical simulations to investigate the low energy states and the emergence of topological defects in an artificial colloidal ice in the Cairo geometry. This type of geometry is characterized by a mixed coordination ($z$), with 
coexistence of both $z=3$ and $z=4$ 
vertices. We realize this particle ice by confining field tunable paramagnetic colloidal particles within a lattice of topographic double wells at a one to one filling using optical tweezers. 
By raising the interaction strength via an applied magnetic field, we find that the ice rule breaks down, and positive monopoles with charge $q=+2$ accumulate
in the $z = 4$ vertices and are screened by negative ones ($q=-1$) in the $z = 3$. The resulting, strongly coupled state remains disordered.
Further, via analysis of the mean chirality associated to each pentagonal plaquette, we find that the disordered ensemble 
for this geometry is massively
degenerate and it corresponds to a frustrated antiferrotoroid.
\end{abstract}

\section{Introduction}
The Cairo geometry is a type of Euclidean plane tiling made of a sequence of connected pentagons 
which share vertices with two types of coordination numbers, namely $z=3$ and $z=4$~\cite{Keeffe1980,Shen2022}.
Besides its aesthetic beauty, as testified by the presence of the Cairo geometry in numerous artistic paintings and pavements, especially in Egypt~\cite{Macmillan1979}, such lattice is also 
important in frustrated spin systems. For example, it has been 
experimentally found in different magnetic compounds 
such as the Bi$_2$Fe$_4$O$_9$~\cite{Ressouche2009} and
the Bi$_4$Fe$_5$O$_{13}$F~\cite{Abakumov2013},
apart from being the subject of theoretical studies on Ising-type models~\cite{Urumov2002,Ralko2011,Rojas2012,Rousochatzakis2012}.
Recently, such geometry has been considered as an interesting way to 
organize interacting dipolar nanoislands on a plane,
also known as artificial spin ice systems (ASIs)~\cite{wang_artificial_2006a,Nisoli2013,Skjarvo2020}.
ASIs are lattice of ferromagnetic elements 
that interact via in-plane dipoles
and are arranged to produce geometric frustration effects~\cite{Mol2010,Rougemaille2011,Zhang2013,gilbert2014emergent,Perrin2016,canals2016,Vedmedenko2016,Wang2016b}.
In the Cairo geometry, recent experimental works have found  
a rich behavior due to frustration~\cite{Saccone2019,Makarova2021}, 
while Monte Carlo simulations reported the presence of long-range 
order~\cite{Shevchenko2022}. Even mechanical analogues of Cairo artificial spin ice were realized via
3D-printing~\cite{Merrigan2022}.

Particle ice systems are soft matter analogues 
of ASIs but based on interacting colloids 
constrained to moved within a lattice of double wells~\cite{Libal2006,Ortiz2019}.
In contrast to ASIs that feature in-plane dipoles, the colloidal particles  present out-of-plane, induced dipoles and pair interactions 
that can be tuned by an external field. The microscopic size of the particles allows the use of optical microscopy to visualize their dynamics and thus, to extract all 
relevant degrees of freedom. The particle ice was originally proposed with a set of
double wells generated optically~\cite{Libal2006,Libal2012}, 
while experiments were realized by using lithographically patterned substrates~\cite{ortiz-ambriz_engineering_2016,Loehr2016,oguz_topology_2020}. 
Moreover, it was shown theoretically that, for a lattice of single coordination number $z$, the 
colloidal ice is analogous to an ASI since the low energy states fulfill similar ice-rules, i.e. minimization of the associated topological charge~\cite{Nisoli14}.
Such similarity however, breaks down for lattices of mixed coordination 
such as decimated systems~\cite{Libal2018}.
Indeed, in a colloidal ice, particles at a vertex tend to repel each other, and therefore the single vertex energy is different from ASI.
If we consider
the colloid as a token of a topological charge, then each vertex wants to push away as
much charge as possible. In certain geometries this is impossible and the same trade-off,
corresponding to the ice rule, is realized on each vertex. Thus, the colloidal ice for an extended, single coordinated lattice behaves as a
spin ice~\cite{Nisoli14,nisoli_unexpected_2018}. In other cases that is not true~\cite{Libal2018}, as we show below for the Cairo geometry, and
a transfer of topological charge takes place between vertices of different coordination, breaking the ice rule.

This fact underlines that particle based ice 
offers the possibility of investigating a rich set of physical phenomena, 
different than ASIs~\cite{Libal2017,nisoli_unexpected_2018,oguz_topology_2020,Gallo2021}. 
And indeed many other recent realizations testify to the broader
phenomenology of particle-based systems as models for geometric frustration. These include confined microgel particles~\cite{Han2008,Shokef2011,Shokef2013,Zhou2017,Leoni2017}, 
mechanical metamaterials~\cite{Kang2013,Coulais2016,Meeussen2020,Shucong2021,Merrigan2021,Meng2022},
patterned superconductors~\cite{Lib2009,Latimer2013}, skyrmions 
in liquid crystals~\cite{Ma2016,Duzgun2021}
and interacting macroscopic rods~\cite{Mellado2012,Mellado2023}. 

\begin{figure*}[t]
\begin{center}
\includegraphics[width=\textwidth]{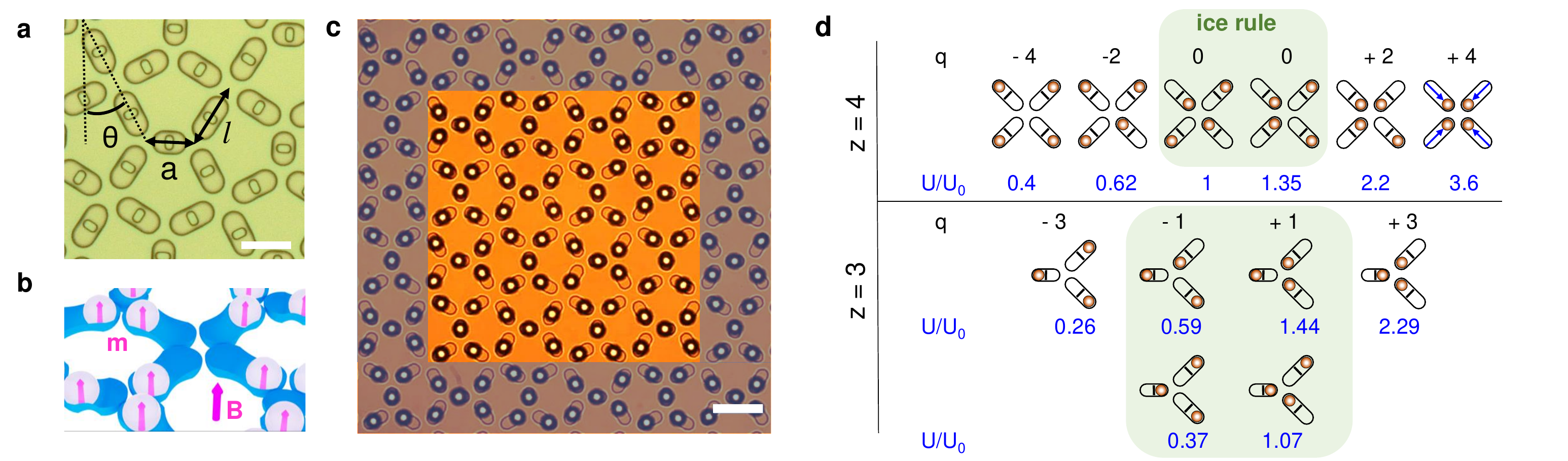}
\caption{\label{figure1}(a) Image 
of a lithographic structure made of 
double wells and arranged along the Cairo geometry with  
the different parameters overlaid: $\theta$ is the bond
angle, $a$ is the distance between two $z = 3$ vertices, $l$ between $z = 3$ and $z = 4$ vertices; scale bar is $10 \, \rm{\mu m}$.
(b) Schematic showing the double wells filled with paramagnetic colloidal particles. The external field $\bm{B}$ induces an equal dipole moment $\bm{m}$ in each particle. 
(c) Experimental realization of a Cairo colloidal ice, where the colloidal particles appear as black disks. The shaded gray region excludes vertices from the statistical analysis, in order to minimize effects
due to open boundaries. Scale bar is 
$20 \, \rm{\mu m}$. (d) Different configurations with corresponding topological charges $q$ (black) and rescaled total energy $U/U_0$ (blue) 
for vertices of coordination $z=4$ (top row) and $z=3$ (bottom row). 
Here $U_0 = 6.2 \rm{k_B T}$ is the energy of the ground state vertex in the $z=4$ for a field of $B=10$ mT.
The last vertex in the $z = 4$ illustrates the Ising-like spins associated to each double well.  The green box shows the vertices obeying the 
ice rules for the coordination $z=4$ ($q=0$)
and $z=3$ ($q=\pm 1$).}
\end{center}
\end{figure*}

In this article we experimentally realize a 
Cairo colloidal ice by confining repulsive paramagnetic colloidal particles within a lattice of lithographic  
double wells, as shown in Fig.~\ref{figure1}(a). 
To place and move these particles within the topographic traps,  
we use a modified set of laser tweezers that generate an optical ring
rather than a focalized spot. This strategy allows to easily trap and move paramagnetic particles, avoiding heating due to absorbed light. By investigating the low energy states, 
we find that topological charges can accumulate in different sublattices, breaking locally the ice rules. We complement our finding with Brownian dynamic simulations, which 
show good agreement with the experimental data in terms of fraction of vertices,
topological charges and net chirality associated to each pentagonal plaquette of the Cairo lattice. Finally we perform simulations on an extended system to 
calculate chirality correlation functions and to elucidate 
the system frustration at high field strength. 

\section{The artificial colloidal ice}
The schematic in Fig.~\ref{figure1}(b) and the optical microscope image in Fig.~\ref{figure1}(c) illustrates the basic features of a Cairo colloidal ice. The system presents a lattice of lithographic elliptical traps composed of two wells of lateral elevation $h \sim 3\rm{\mu m}$
that are connected by a small central hill. These wells are filled with paramagnetic colloidal particles of 
diameter $d=10 \, \rm{\mu m}$ and magnetic volume susceptibility $\kappa=0.025$. 
A particle of volume $V=\pi d^3/6$ has to overcome a gravitational potential $U_g=\Delta \rho V g h \sim 2000 \, \rm{k_B T}$ to jump outside the double well due to thermal fluctuations. Here $\Delta \rho = \rho_p - \rho_w \sim 0.6 \, \rm{g \, cm^{-3}}$  is the difference between the mass density of the particle ($\rho_p$) and the dispersing medium ($\rho_w$). 
Thus, the particles are essentially confined within the elliptical traps and cannot change their location from one well to another unless subjected to an external force, such as the repulsion from a neighboring colloid.  We induce such repulsion by applying an external magnetic field $\bm{B}$ perpendicular to the sample pane, Fig.~\ref{figure1}(b). 
Once the field is applied, each particle acquires an induced dipole moment $\bm{m}=V \kappa \bm{B}/\mu_0$,  
being $\mu_0$ the permeability of the medium. 
Thus, a pair of particles $(i,j)$ placed
at a relative distance $r=|\bm{r}_i-\bm{r}_j|$
experience a repulsive dipolar interaction 
which is isotropic and inversely proportional to 
$r^3$, $U_{dd}= \mu_0 m^2/(4\pi r^3)$.
This interaction potential can be tuned via the applied field and, for an amplitude of $B =10$mT, $U_{dd}= 122 \, k_B T$ ($U_{dd}= 4.7 \, k_B T$) for the closest (farthest) distance of $r= 13\rm{\mu m}$ between two particles in a $z=3$ vertex ($r= 38.4\rm{\mu m}$ in a $z=4$ resp.).

The mapping between the colloidal ice and an ASI~\cite{Libal2006}
can be obtained by assigning an Ising-like spin to each double well,
such that it points where the particle is located Fig.~\ref{figure1}(d). 
Using this mapping, one can distinguish between different type of vertices
depending on the lattice coordination. For example, for the $z=4$ (square lattice)
there are $6$ possible arrangements of the particles with different energetic weights,
while for the $z=3$ (honeycomb lattice) these reduce to $4$.
Moreover, in analogy to the ASI, one can assign a topological charge to each 
vertex defined here as $q=2n-z$ being $n$ the number
of spins that point toward the vertex center.
Note that we can talk of topological charges when considering the 
vertices within a lattice, not isolated ones.
By this notion of charge, an extended lattice is overall charge-neutral, and thus charges appear in pair and disappear only when annihilated by other
defects in order to guarantee the charge conservation, $\sum q_{i} =0$. For example, the 
vertex with four colloids pointing outwards ($q=-4$) 
is characterized by the lowest energetic weight of the $z=4$, 
and thus, when considered alone it will be the natural state of repulsive colloids. However, within a lattice
the $q=-4$ is topologically connected to the $q=+4$ 
due to particle conservation and thus they are unlikely to occur.
One can prove that in a lattice, lower absolute values of the topological charges $|q|=0,1$,
corresponding to the ice-rule, are favored ~\cite{nisoli_unexpected_2018}.
Indeed the ice rules, highlighted by the green box in Figure~\ref{figure1}(d) 
are a prescription of the minimization of $|q|$,
given by vertices with $q=0$ for $z=4$
and $q=\pm 1$ for $z=3$.

Regarding the Cairo geometry, we have a mixture of $z=4$ and $z=3$ vertices, the latter characterized by unequal 
lengths of the double wells, as shown in Fig.~\ref{figure1}(d).  Upon analysis of the total magnetic interaction energy of a vertex $U$, 
we find that the $z=4$ vertices 
have same energy hierarchy 
as the square colloidal ice ice investigated in previous works~\cite{ortiz-ambriz_engineering_2016,Loehr2016}. 
Here $U=\sum_{i=1}^{N_v-1}\sum_{j=i+1}^{N_v}U^{dd}_{ij}$, being $N_v$ the number of particles in a vertex. 
However, the presence of the small 
double well in the $z=3$ vertices, 
i.e. a length of $4.53 \, \rm{\mu m}$ while in the $z=4$ is $10 \, \rm{\mu m}$,
induces an energetic spitting of the
vertices $1$-in-$2$-out and $2$-in-$1$-out depending on the location of the paramagnetic colloid.
This energy difference between the $z=3$ vertices,  which does not affect the associated topological charge,   is 
particular of the Cairo geometry, and it is not present in the $z=3$ vertices of the classical honeycomb~\cite{Lib2018,Cunuder2019} and triangular~\cite{Dong2018} colloidal ice, where all traps have the same length. 
Even if the energy difference is relatively small, as we will show later this will induce a disordered ground state.

\begin{figure}[ht]
\begin{center}
\includegraphics[width=0.8\columnwidth]{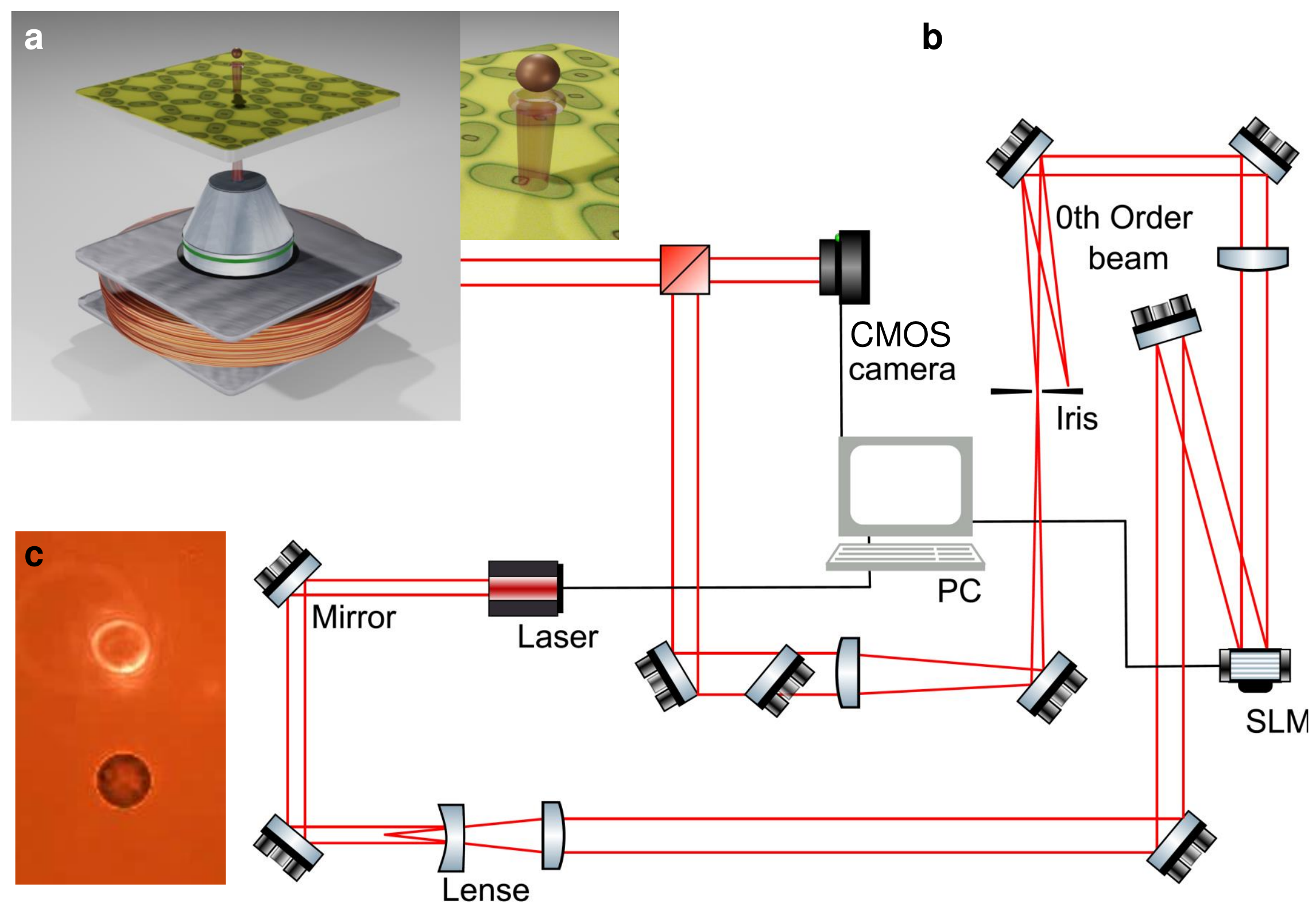}
\caption{\label{figure2}(a) Schematic showing a colloidal particle
trapped above the topographic substrate by laser tweezers. The beam is focused with a microscope objective that passes through
a magnetic coil. Side image shows an enlargement of the particle and 
the optical ring. (b) Detailed sketch of the experimental system illustrating the different optical components to 
send a near-infrared laser beam onto a spatial light modulator (SLM), and the system dynamics are visualized via a complementary metal oxide semiconductor
(CMOS) camera. 
(c) Microscope image of a particle and the corresponding optical ring.}
\end{center}
\end{figure}

\section{Experimental methods}
The Cairo lattice is realized via soft-lithography using 
Polydimethylsiloxane (PDMS). This substrate was first designed using a commercial software  (AutoCad, Adobe) and fabricated above a $5$-inch soda-lime glass
covered with a $500$nm
Cromium (Cr) layer. 
The chosen geometric parameters, as shown in Fig.~\ref{figure1}(a), are 
$a=19.54 \, \rm{\mu m}$, $l=26.7 \, \rm{\mu m}$ and $\theta=30^{\circ}$.
We use laser lithography (DWL 66, Heidelberg
Instruments Mikrotechnik GmbH)
to write the double wells on the substrate with a $405$nm laser
diode working at a speed of $5.7 \rm{\, mm^2 \, min^{-1}}$. 
After that, we replicate the double wells of the  Cr mask in the PDMS following two steps. In the first one we duplicate the structure using 
an epoxy-based negative photoresist (SU-8) 
on top of a silicon wafer.
Then we covered the structure with liquid PDMS 
by spinning the sample at $4000$ rpm during $1$ min with an angular speed of $2000$ rpm (Spinner Ws-650Sz, Laurell). With this process we obtain 
a layer of $\approx$ 20 $\mu m$ thickness. 
The PDMS is solidified by baking for $30$ min at $95^{\circ}$C in a leveled hot plate.
After solidification of the PDMS, 
we peel off the structure with the help of a cover glass (MENZEL-GLASER, Deckglaser). The resulting sample is  $\sim 170 \, \mu m$ thick, and enough transparent to visible light.

Once fabricated, the sample is placed on the stage of an inverted optical microscope (TiU, Nikon)
which is connected to a complementary metal oxide semiconductor camera (MQ013CG-E2, Ximea) able to record videos of the particle dynamics at $30$Hz.  
The microscope is equipped with a  
$40\times$ oil immersion objective (Nikon, numerical aperture $1.3$) which is 
used both for 
observation and optical trapping purpose.

One microscope port is modified in order to accommodate
an incoming beam 
generated by  
a butterfly laser diode (wavelength $\lambda  = 976$nm, operated at a power of $70$mW, BL976-SAG300 Thorlabs).
The optical path of the laser is composed of a series of optical lenses 
including a spatial light modulator (SLM, Hamamatsu X10468-03) which is commanded by a LCOS-SLM controller (Hamamatsu), Figs.~\ref{figure2}(a,b). The holograms are generated with a custom made Labview program. 

The SLM is used to generate holographic optical tweezers  (HOTs)
which are composed of $4$ lenses. The first two constitute a telescope and have focal lengths $f = -75$ mm (Thorlabs LC 1582-B) and $f = 175$ mm (Thorlabs LA 1229-B), respectively.
After the SLM there is another telescope with a lens of focal $f = 500$ mm (Thorlabs LA1908-B) which focuses the beam after being deflected in order to filter the $0$ mode via a diaphragm, and reaching a final lens of $f = 750$ mm (Thorlabs LA 1727-C), as shown in the detailed diagram of Fig.~\ref{figure2}(b). 
Due
to the optical absorption of the particles used, which are
highly doped with nanoscale iron oxide grains ($\sim 20\%$ by
w.), we have implemented a novel strategy to trap the
colloids without damaging them due to the generated
heat. Instead of using a single focalized spot, we program the SLM such that the deflected
beam generates an optical ring as shown in Figs.~\ref{figure2}(a,c).

The external magnetic field was generated via a custom made coil
located below the sample. The magnetic coil is powered by an amplifier  (BOP-20 10M, KEPCO), that is computer controlled using a digital analogue card (NI 9269) with a custom made LabVIEW program.
The field was applied via a ramp at a rate $0.05 \, \rm{mT s^{-1}}$
until reaching a maximum value of $10$ mT.

\section{Numerical simulation}
We complement the experiments performing Brownian dynamics simulations
using as input parameters
the experimental data. 
In particular, we use Euler's method to integrate 
the overdamped equations of motion for each colloidal particle $i$ at position $\bm{r}_i$: 
\begin{equation}
\gamma\frac{d \bm{r}_i}{dt} = \bm{F}_i^{\mathrm{T}} + \bm{F}_i^{\mathrm{dd}} + \bm{\eta} \, \, \, ,  
\label{equation1}
\end{equation}
being $\gamma = 0.033 \rm{pN \, s \, m^{-1}}$ the friction coefficient.
Further terms in Eq.~\ref{equation1} are the force from 
the topographic double well $\bm{F}_i^{\mathrm{T}}$ which is is modeled as a piece-wise harmonic bistable potential,
\begin{equation}
\bm{F}_i^{\mathrm{T}}=-\bm{e}_{\perp}kR_{\perp}+\bm{e}_{\parallel}\delta \, \, \, .
\end{equation}
Here $(\bm{e}_{\parallel},\bm{e}_{\perp})$
are unit vectors oriented parallel
and perpendicular with respect to the line of length $L$ that joins
the two minima in the double well, whose associated vector is 
$\bm{R}\equiv (R_\parallel,R_\perp)$. Moreover
$\delta=\xi_1 R_\parallel
$ if $|R_\parallel| \leq \frac{\delta}{2}$
and $\delta=k (\frac{L}{2}-|R_\parallel|) \mathrm{sign} (R_\parallel)$ otherwise. 
As stiffness we use $k = 1 \cdot 10^{-4} \, \rm{ pN/nm}$ which keeps the particle confined to the elongated region around the center of the trap, and $\xi_1 = 3  \  \rm{pN/nm}$ that creates a potential hill equivalent to the 
gravitational hill within the double wells. 

The dipolar force on a particle $i$  is given by, 
\begin{equation}
\bm{F}_i^{\mathrm{dd}} = \frac{3\mu_0}{4\pi} \sum_{j\neq i}  \frac{\bm{m}^2\hat{r}_{ij}}{|r_{ij}|^4} \, \, \, ,
\end{equation}
being $\mu_0=4\pi \times 10^{-7} \rm{H/m}$ and $\hat{r}_{ij}=(\vec{r}_i-\vec{r}_j)/|\vec{r}_i-\vec{r}_j|$. Dipolar interactions are calculated in an iterative form such that the global field $\bm{B}$ includes also that generated by all other dipoles. Moreover, we apply a large cut-off of $200 \rm{\mu m}$ to consider the effect of long range dipolar interactions. 

Finally the last term in Eq.~\ref{equation1} is a random force characterized by a zero mean, $\langle \bm{\eta} \rangle = 0$ and delta correlated, 
$\langle \bm{\eta} \left(t\right) \bm{\eta} \left(t'\right) \rangle = 2k_\mathrm{B}T\gamma \delta(t-t')$, being $k_\mathrm{B}$
the Boltzmann constant and
$T=300\mathrm{K}$ the ambient temperature. 

To reproduce the experimental results we start by solving 
Eq.~\ref{equation1} with $N_1 =180$ 
particles which are arranged along $3\times 3$ unit cells
and open boundary conditions, similar to Fig.~\ref{figure1}(c). 
However, to consider 
a larger system when measuring the chirality correlation 
functions, we extend the simulations also to 
$N_2=2000$ particles, where
$800$ are arranged along the $z = 3$ vertices and $400$ in the $z = 4$.
This situation corresponds to a colloidal ice of $10 \times 10$ unit cells also with open boundary conditions. To avoid that most of the particles in the $z=3$ vertices localize on top of the topographic hills, due to strong dipolar forces, we also reduce 
the particle magnetic susceptibility to $\kappa_2 = 0.005$
and raise the spring constant of the central hill to $\xi_2=25    \rm{pN/nm}$. 
In all cases, we numerically integrate the equation of motion using a time step of $\Delta t = 0.01$s.

\section{Measurements of the topological charges.}
\begin{figure}[t]
\begin{center}
\includegraphics[width=\columnwidth]{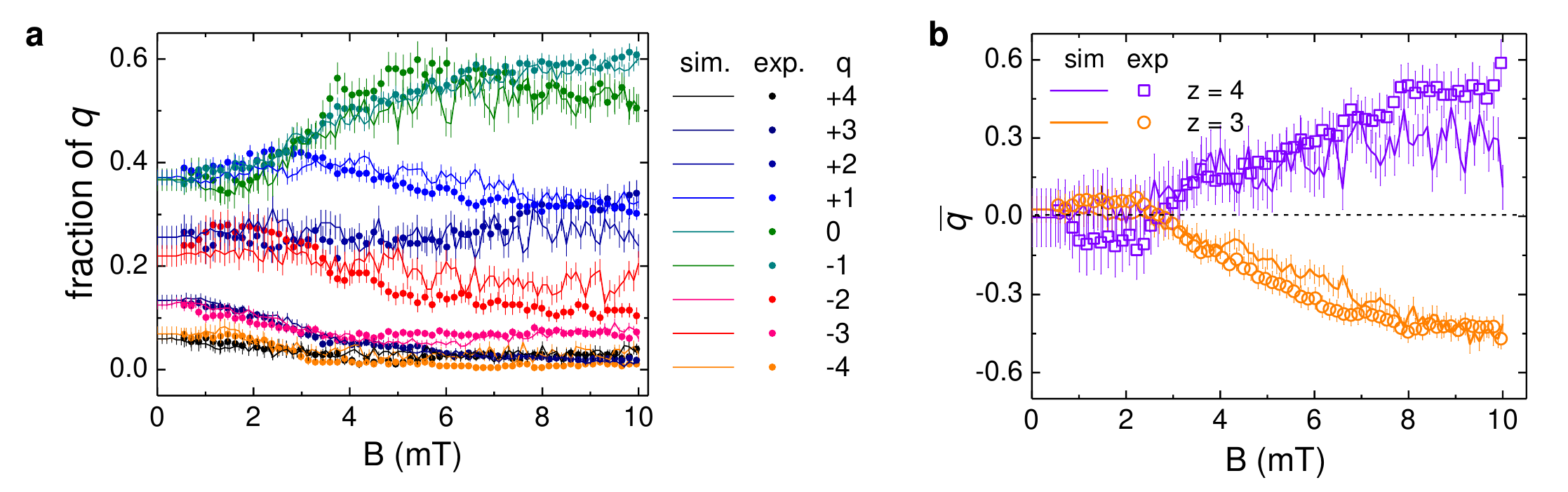}
\caption{\label{figure3}(a)
Fraction of topological charges $q$ for all
vertex types (legend on left), and (b) mean topological charge
$\bar{q}$ for $z = 3$ (empty circles) and $z = 4$ (filled squares) vertices versus applied field strength $B$.
In both graphs the symbols are experimental data, 
while the lines are results from numerical simulations.}
\end{center}
\end{figure}

We start our experiments by first 
randomly placing the particles with the optical ring within the double wells according to a
random number generator.
In the initial, random configuration
the highly charged monopoles
$q=\pm 4$ are the $10\%$ of the total vertices,
$q=\pm 3$ are the $15\%$,
while the $25\%$ of vertices corresponds to low charged
$q=\pm 2$, the rest 
is leave to the ice rule vertices. 
Then, we slowly raise the applied field
with a ramp of $0.05 \, \rm{mT \, s^{-1}}$. 
Fig.~\ref{figure3}(a) shows the evolution of the fraction of topological charges as classified in Fig.~\ref{figure1}(d),
for the Cairo ice. 
By increasing $\bm{B}$, we find that 
already after $B\sim 3$ mT, the fraction of 
high topological charges $q=\pm 4$ in the $z=4$
and $q=\pm 3$ in the $z=3$
vertices reduces almost to zero
in favor of the low charged ones. 
In particular, vertices obeying the ice selection rules
in the $z=4$ rise up to the $50\%$,
being only overcomed by the $q=-1$
in the $z=3$ ($\sim 60\%$),
while the $q=+1$
reduces to  $\sim 50\%$.
This reduction is 
accompanied by a slight increase of the
charged monopoles $q=+2$
and a decrease of the $q=-2$.

Crucially, Fig.~\ref{figure3}(b) plots the average topological charge ($\bar{q}=\frac{1}{N_z}\sum q_{z_i}$
$N_z$ being the number of vertices $z$) for vertices of coordination $3$ and $4$. It shows a net separation of
topological charges between vertices of different coordination, and thus charge transfer
between sublattices, breaking the ice rule. Thus,
while the total topological charge is conserved,
or $\sum q=0$,
at the sublattice  level we observe 
a transfer of topological charges 
as the field reaches $B=8$ mT.
A net fraction of positive monopoles
accumulate in the $z = 4$ vertices, while the 
$z = 3$ vertices are, on average, negatively charged.
Note also the relatively good agreement between experimental data 
(open symbols) 
and numerical simulations (continuous lines), which are plot together in both graphs, while small deviations fall within the experimental/simulation error bars.

Importantly, charge transfer effect between sublattices does not occur in an ASI, whose ice rule is instead  robust in most geometries
as it is inscribed into the energetics of the vertices. It only occurs in a colloidal ice with 
mixed coordination geometry as the Cairo.
This effect results from the different nature of geometric frustration in ASI and in the colloidal system~\cite{Nisoli14}.
While both systems display similar behavior in terms of vertex fraction for single coordination lattices, 
in a mixed coordination geometry the difference at the single vertex level emerges: repulsive colloids cannot emulate 
in-plane ferromagnetic spins as in ASI, and topological charges tend to redistribute in order to minimize the system energy. 

While charge transfer and ice-rule fragility had been predicted~\cite{Nisoli14} and
experimentally verified~\cite{Libal2018} in a decimated square ice, the sign of the effect in the Cairo lattice
is inverted. In Ref.~\cite{Libal2018}, negative monopoles form on the $z = 4$, breaking the ice rule.
In that case the $z = 3$ vertices, which unlike the $z = 4$ vertices are charged even when
obeying the ice rule because of their odd coordination, do not violate the ice rule but
screen the negative charge of the $z = 4$ vertices by changing their relative admixture of
$\pm1$ charges, and thus assuming a net positive charge. In the Cairo system instead, the mechanism is similar
but the sign of the charges is inverted. This is because Cairo is structurally different from a decimated square, with shorter and longer traps and thus a more complex energetics, as shown in Fig.~\ref{figure1}(d).
This allows also for structural transitions in the sign of the transferred charge as Cairo is
deformed, and which we study elsewhere, while
here we focus on the Cairo geometry.

\begin{figure*}[t]
\begin{center}
\includegraphics[width = 0.8\textwidth]{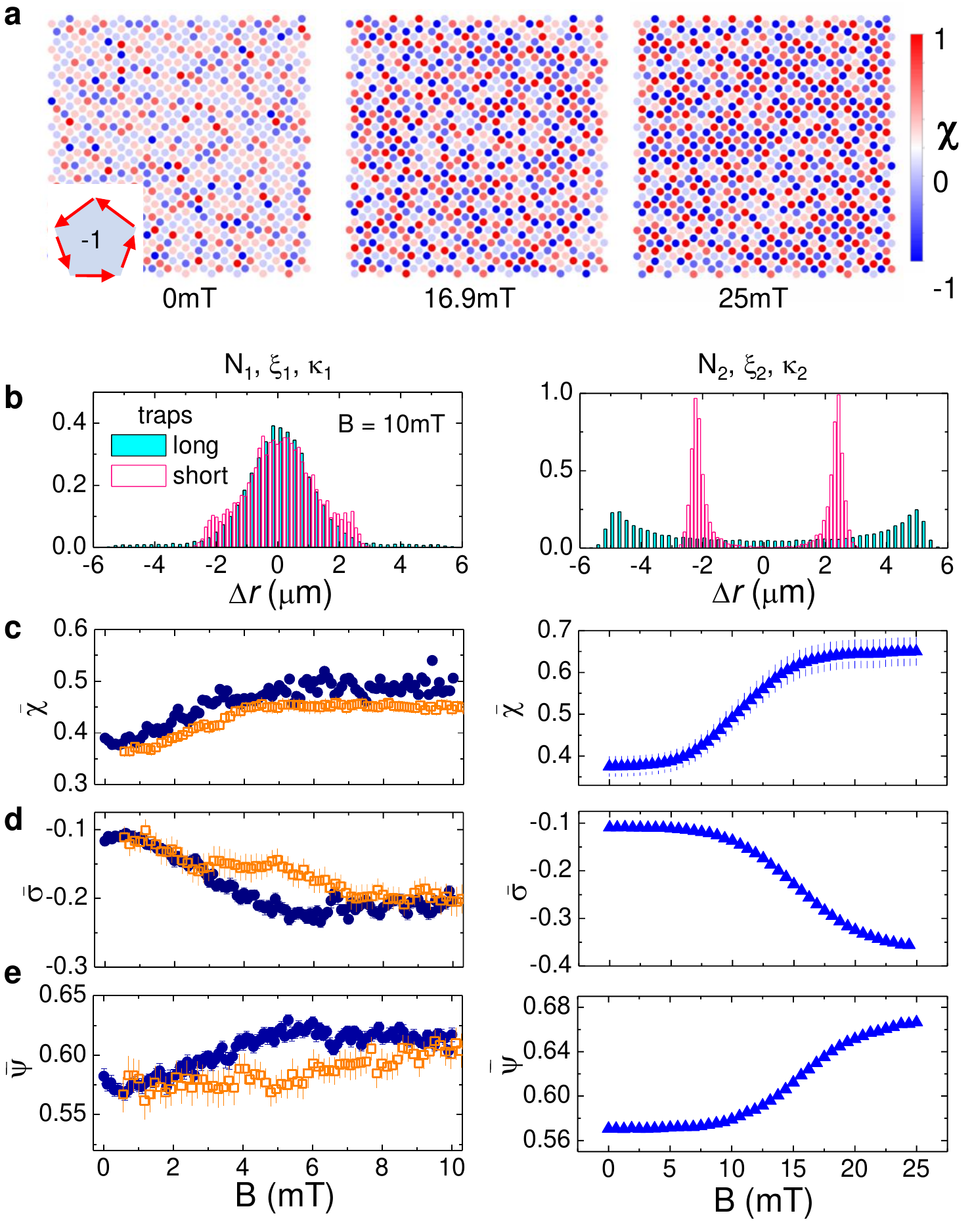}
\caption{(a) Colormaps at different field amplitudes  showing the evolution of the chirality $\chi$ associated to each pentagonal plaquette for a system of $100\time 100$ plaquettes. 
Small inset at the bottom of the first image illustrates a counter clockwise plaquette with $\chi=-1$. 
(b-e) Different panels correspond to two system size
and parameters for simulations: left column $N_1=180$ particles, 
$\xi_1=3 \, \rm{pN \, nm^{-1}}$
and $\kappa_1= 0.025$
right column $N_2=2000$ particles, 
$\xi_2=25 \, \rm{pN \, nm^{-1}}$
and $\kappa_2= 0.005$.
In the first row (b) are the
Histograms of positions of colloidal particles in the long (cyan) and short (pink) traps.
Second (c) the mean chirality $\bar{\chi}$,
third (d) the correlation $\bar{sigma}$ (d) and 
fourth (e) row the correlation $\bar{\Psi}$
all versus applied magnetic field $B$.
In all graphs on the left column, orange squares 
indicate experimental data 
while navy disk simulation one 
obtained for a $10 \time 10$ vertex lattice.
Blue triangles on the right graphs are  
simulation results for  
a $100 \time 100$ vertex lattice.}
\label{figure4}
\end{center}
\end{figure*}

\subsection{Frustrated antiferrotoroidicity in the Cairo system}

To better characterize the disorder of the charge-unbalanced, low-energy state of
strongly-interacting Cairo colloidal ice, we study the chirality value, or associated toroidal moment, to each pentagonal plaquette. As shown in the inset in Fig.~\ref{figure4}(a), this moment 
acquires a maximum value of  $\chi = -1$ ($+1$) for a full counter clockwise (clockwise) cell. Two natural questions are, firstly whether the plaquettes are
naturally chiral, and secondly what is the mutual arrangement of these chiralities.
To answer the first question, note that for a single pentagonal plaquette, a configuration that is fully chiral obeys the ice rule,
that we know breaks down at high fields.
At the same time, the long range interactions among colloids in further neighboring traps favors chirality. In the Cairo geometry the plaquette chirality is also favored by the asymmetry in the vertices of coordination $z=3$. 
Because the traps between two of these vertices
are considerably shorter than the others, moving these colloids \textit{in}, i.e. towards the vertex center, it leads to a smaller change in energy than moving colloids along the longer traps. Thus, the particles in the two longer traps prefer to stay head-to-toe which gives a contribution $2/5$ to chirality. However, those longer traps impinge on vertices of coordination $z=4$. Assuming that they preferentially are in an antiferromagnetic configuration, as shown in the third image in Fig~\ref{figure4}(a), that adds another $2/5$ contribution to the chirality for a total of $\chi=0.8$.

Thus we expect that at large field strengths, 
the average absolute chirality of the systems,
$\bar{\chi}=\frac{1}{N_{pl}}\sum_{i}|\chi_i|$ 
being $N_{pl}$ the number of plaquette,
would tend towards $\bar{\chi} \rightarrow 0.8$.
However we found that this was not the case, as shown in Fig~\ref{figure4}(c), where both experiments and simulations show that at the largest applied field $\bar{\chi} \sim 0.5$. 
This lower value was due to the fact that,
in the Cairo geometry, the magnetic colloids were found to locate above the 
central hill at the largest field amplitude, $B=10$mT.
Indeed this effect is shown in the first graph in Fig~\ref{figure4}(b),
where histograms of the particle positions from the simulations are reported for both the long and short 
double wells in the Cairo geometry.  In this situation, since the trap bistability is lost, it is difficult to extract an accurate determination of $\bar{\chi}$. 

To circumvent this problem numerically, 
we have performed further simulations using a larger system size and stronger confinement, i.e. increasing the hill spring constant from $\xi_1 \to \xi_2=25 \, \rm{pN \, nm^{-1}}$
and decreasing the magnetic volume susceptibility to $\kappa_1 \to \kappa_2=0.005$.
The resulting histograms of the particle positions, shown in the right panel of Fig.~\ref{figure4}(b),
confirm that for these new parameters the bistability is recovered
even for a larger field of $B=25$mT.
Thus,  for the latter system, we obtain the value of $\bar{\chi} \sim 0.7$
at the largest field which is in the ballpark of our estimate  $\bar{\chi} = 0.8$ based on a single plaquette. Note that Fig.~\ref{figure4}(a) shows an alternation of sign among many nearest neighboring plaquettes. However, the lattice of the pentagonal plaquettes is not bipartite, and thus frustrates the anti alignment of the plaquettes. A configuration in which all the neighboring plaquettes have opposite chirality (e.g., the
check board pattern observed for the square spin ice system~\cite{Morgan2011}) is geometrically impossible which suggests that the
system has, at least, a disordered landscape of low energy
states. As Figure~\ref{figure4}(a) suggests, the configuration corresponds to an antiferrotoroid. 
Moreover, we have check that also with the new numerical conditions ($N_2,\xi_2,\kappa_2$)
we observe the topological charge transfer 
similar to Fig.~\ref{figure3}(b).

To answer the second question, concerning the mutual distribution of chiralities, we
note that Fig. 4(a) shows a largely antiferrotoroidal arrangement. However, the lattice
of the plaquettes is frustrated and therefore no full antiferrotoroidicity can exist. We
explore this angle by extracting the following nearest neighbor correlations:
\begin{equation} 
\bar{\psi}= \frac{1}{N_{nn}}\sum_{\langle ij \rangle}(1-\frac{\chi_i \chi_j}{|\chi_i \chi_j|})\frac{1}{2}\, \, ; \, \, \, \, \, \, \bar{\sigma}= \frac{1}{N_{nn}}\sum_{\langle ij \rangle}\frac{\chi_i \chi_j}{\bar{\chi}} \, \, ;
\end{equation} 
where the sum is performed over 
the $N_{nn}$ nearest neighboring plaquettes. Both correlations counts how many links among nearest neighboring
plaquettes are antiferrotoroidal, regardless of the intensity of the chiralities of the
plaquettes. If all the nearest neighboring plaquettes were antiferrotoroidal (which is
impossible because of the frustration of the lattice of the plaquettes) then $\bar{\psi}$ would be equal to $1$,
and $\bar{\sigma} \to -1$.

We plot $\bar{\psi}$ and $\bar{\sigma}$ in Figs.~\ref{figure4}(d,e).
In the left images, both correlations show similar trends 
between experiments and numerical simulations. 
For the large system size (right images) $\bar{\psi}$ increases steadily with the field reaching $0.68$: almost the $68\%$ of the links are anti-ferromagnetic, namely $3.5$ over $5$ nearest neighbors. This results also reflects the arrangement of the toroidal moments shown in Fig.~\ref{figure4}(a) for $B=25$ mT, where the system organizes in a lattice of full chiral cells placed in an alternating order.  These results allow to characterize the disorder of
the strongly coupled state of the Cairo ice in terms of plaquette chirality as a frustrated
antiferrotoroid. The charge transfer prevents all plaquettes from reaching maximum
chirality, while the the lattice of the plaquettes frustrates antiferrotoroidicity, preventing order.

\section{Conclusions}
We have investigated the arrangement of repulsive magnetic colloids confined in a lattice of double wells in the Cairo geometry. 
To experimentally realize such structure we have modified the 
lithographic process and developed a novel technique to trap absorbing magnetic particles using 
an optical ring.
This strategy could be extended to other
works aiming at trapping light-adsorbing magnetic particles 
to avoid undesired heating effects. 
We have observed that the Cairo ice 
breaks the ice rule as predicted for mixed coordination geometries, however, it does so with an inversion of the net charge transfer with 
respect to a previous experimental realization~\cite{Libal2018}.
Moreover we have characterized this novel
ensemble but looking at the effective toroidal moment associated to each 
pentagonal plaquette. We have found that the strongly coupled ensembled of the Cairo colloidal ice is a  massively degenerate  antiferroid. 

Further extensions of our work include investigating the transition from Shakti to Cairo (ongoing)
or using different  size of particles to investigate hysteresis and memory
effects~\cite{Libal2012,Gallo2021} that could emerge when the particles localize above the double wells~\cite{Gallo20212}.
Finally, it will be also interesting to investigate how the topological charges freeze or move after long time and the presence of 
aging of topological defects in our system. This will require longer observation time, beyond our current experimental capabilities, and thus could be a challenge for future work. 

This project has received funding from the 
European Research Council (ERC) under the European Union's Horizon 2020 research and innovation programme (grant agreement no. 811234).
P. T. acknowledge support the Generalitat de Catalunya under Program ``ICREA Acad\`emia''. The work of Cristiano Nisoli was carried out under the
auspices of the U.S. DoE through the Los Alamos National Laboratory, operated by
Triad National Security, LLC (Contract No. 892333218NCA000001).

\bibliographystyle{iopart-num}
\providecommand{\newblock}{}

\end{document}